\documentclass[reprint,unsortedaddress,showkeys,amsmath,amssymb,aps,showpacs]{revtex4-1}
\usepackage{graphicx}
\usepackage{dcolumn}
\usepackage{bm}
\usepackage[colorlinks,linkcolor=blue,citecolor=blue]{hyperref} 
\usepackage{epstopdf}

\def\pr{\prime}
\def\be{\begin{equation}}
\def\lan{\left\langle}
\def\ran{\right\rangle}
\def\ee{\end{equation}}
\def\barr{\begin{array}}
\def\earr{\end{array}}

\def\nn8{\\}
\def\l{\left}
\def\r{\right}
\def\dis{\displaystyle}
\def\ed{\end{document}}

\def\cx{{\cal X}}

\def\cx{{\cal X}}

\def\cq{{\cal Q}}

\def\cas{{\cal S}}
\def\car{{\cal R}}
\def\capp{{\cal P}}

\def\cs{{\bf s}}

\def\spin{\frac{1}{2}}

\begin{document}

\title[Fidelity decay and entropy production]{Fidelity decay and entropy
production in many-particle systems after random interaction quench}

\author{Sudip Kumar Haldar}
\affiliation{Theoretical Physics Division, Physical Research Laboratory,
Navarangpura, Ahmedabad 380009, India.}
\email{sudip@prl.res.in}

\author{N.D. Chavda} 
\affiliation{Applied Physics Department, Faculty of Technology and 
Engineering, Maharaja Sayajirao University of Baroda, Vadodara 390 001, India.}
\email{ndchavda-apphy@msubaroda.ac.in}

\author{Manan Vyas}
\affiliation{Instituto de Ciencias F\'{i}sicas, Universidad Nacional 
Aut{\'o}noma de M\'{e}xico, C.P. 62210 Cuernavaca, M\'{e}xico}
\email{manan@icf.unam.mx}
\thanks{Corresponding author}

\author{V.K.B. Kota}
\affiliation{Theoretical Physics Division, Physical Research Laboratory,
Navarangpura, Ahmedabad 380009, India.}
\email{vkbkota@prl.res.in}

\date{\today}

\begin{abstract}

We analyze the effect of spin degree of freedom on fidelity decay and entropy
production of a many-particle fermionic(bosonic) system in a mean-field,
quenched by a random two-body interaction preserving many-particle spin $S$.
The system Hamiltonian is represented by embedded Gaussian orthogonal 
ensemble (EGOE) of random matrices (for time-reversal and rotationally invariant 
systems) with one plus two-body interactions preserving $S$ for fermions/bosons.
EGOE are paradigmatic models to study the dynamical transition from integrability to chaos in interacting many-body quantum systems. A simple general picture, in which the variances of the eigenvalue density play a central role, is obtained for describing the short-time dynamics of fidelity decay and entropy production. Using some approximations, an EGOE formula for the time ($t_{sat}$) for the onset of saturation of entropy, is also derived. These analytical EGOE results are in good agreement with numerical calculations. Moreover, both fermion and boson systems show significant spin dependence on the relaxation dynamics of the fidelity and entropy. 

\end{abstract}

\pacs{05.30.-d, 05.70.Ln, 05.45.Mt, 05.30.Fk, 05.30.Jp}

\keywords{statistical relaxation, thermalization, random interaction quench, 
embedded random matrix ensembles, quantum many-particle systems}

\maketitle

\section{Introduction}

Investigation of non-equilibrium dynamics and statistical relaxation in isolated,
interacting, many-body quantum systems has emerged as an important research
area in the last few decades~\cite{Sengupta, science-review, Rigol-rev}. 
Magnificent experimental progress, particularly in the field of ultracold atoms,
has made it possible to artificially realize interacting many-body quantum systems to understand various quantum phenomenon. Systematic understanding of non-equilibrium dynamics of such complex systems is still not available. This is not only a fundamental requirement but may also be crucial for applications in quantum information \cite{Sengupta}. A complex many-particle quantum system exhibits non-equilibrium dynamics following a change in one of the system parameters, usually referred to as quantum quench. Coherent quench dynamics has been observed for both bosonic~\cite{cqd-b} and fermionic~\cite{cqd-f} systems. The main questions addressed in these studies are to understand: (i) whether and how a quantum system thermalizes, (ii) universal aspects of the dynamics, (iii) the relation between localization, integrability and thermalization, and (iv) behavior of various observables after equilibration. Although theoretical advancements have been made to understand these questions using spin lattice models~\cite{Rigol-4, Eck-09, Rigol-2, Wright, Rigol-3, Manan, Cal-06, Cal-13, Fag-13, Poz-14, Pro-15}, deriving universal properties for such complex systems remains an open question. 

It is now generally accepted that statistical relaxation in the quench dynamics of an isolated interacting many-body quantum system is closely related to quantum chaos \cite{ETH, Rigol2008, Lea-10}. Signatures of quantum chaos appear in the statistical behavior of eigenvalues and eigenfunctions \cite{Haake}. Quantum integrable systems have been found to relax to an equilibrium state 
characterized by the generalized Gibbs ensemble~\cite{Rigol-3, Caux, Cal-06, Rigol-11, Cal-13, Fag-13, Poz-14, Pro-15}. On the other hand, non-integrable quantum systems relax towards thermal equilibrium. Therefore, for these systems, for example, expectation values of various few-body observables approach their long time average values given by the Gibbs ensemble. The eigenstate thermalization hypothesis (ETH) is considered to be the underlying mechanism for thermalization in isolated quantum systems~\cite{ETH, Rigol2008}. ETH states that the eigenstate expectation values of typical observables are smooth functions of energy eigenvalues with off-diagonal elements being exponentially small in system size. Various aspects related to ETH have been studied by many groups using lattice models of interacting spins, fermionic as well as bosonic; see for example ~\cite{Lea-12, Kollath, ETH-1, ETH-2, Iz-PRL, Manan}. 

In order to understand the role played by quantum chaos in attaining thermalization of isolated finite quantum systems, thermodynamical description of isolated interacting quantum systems using random interactions was first addressed in \cite{Fl-97}. EGOE are paradigmatic models to study many-body quantum chaos \cite{PW-07}. The constituents of real physical systems interact predominantly via two-body interactions in the presence of an average one-body (mean-field) interaction. Flambaum and Izrailev started investigation of time evolution of generic quantum many-body systems~\cite{Fl-Aust, FlIz}. Going further, EGOE(1+2) generated by one plus two-body interactions was employed in \cite{KoReta} to investigate thermalization in isolated quantum systems using ergodicity principle. One of the most significant aspects of EGOE(1+2) is that the system undergoes a transition from regular to chaotic region with increasing strength of the random two-body interaction quench that deeply affects eigenvalue and eigenfunction structure \cite{Ko-14}. These random matrix ensembles are used so far to study (i) ergodicity principle for expectation values of several different operators(observables); (ii) determine a region of thermalization using the criterion of equivalence between different definitions of entropy and temperature; (iii) representability of occupancies of single particle states using Fermi-Dirac (or Bose-Einstein) distribution; (iv) calculation of expectation values using canonical distribution. See \cite{KoReta, Ko-01, Go-12} and references therein. It was also shown that the onset of Wigner's surmise in spectral fluctuations is not sufficient for finite systems to thermalize \cite{KoReta}.

EGOE are generic, though analytically difficult to deal with, compared to lattice spin models as the latter are associated with spatial coordinates (nearest and next-nearest neighbor interactions) only. Moreover, the universal properties derived using EGOE extend easily to systems represented by lattice spin models, for example eigenvalue density is Gaussian in the mean-field basis unlike a classical GOE \cite{Manan}.  Due to two-body character of the interactions, the three different types of non-zero many-particle matrix elements (zero, one and two particle transfer) can be derived analytically by action of creation and annihilation operators on the many-particle states in the Fock-space representation with the system Hamiltonian (see Eq. (\ref{eq-1}) ahead) expressed in second quantized form. These non-zero matrix elements will be a linear combination of two-body matrix elements (chosen to be GOE) and hence, correlated. Due to this, matrix structure for an EGOE is different from a classical GOE which accounts for many-body interactions among system constituents (although fluctuation properties of EGOE are same as that of a GOE) \cite{Man-10, Ko-14}.

In addition to particle number, spin $(S)$ quantum number is important for realistic systems like atomic nuclei, quantum dots, nano-metallic grains, ultracold spinor gases etc. As group symmetries define various quantum numbers, in general, one has to consider EGOE with group symmetries. The first non-trivial and important (from the point of view of its applications) embedded ensembles are EGOE(1+2)-$S$ [BEGOE(1+2)-$F$] with spin degree of freedom for fermions [bosons] \cite{Ma-PRE, Ma-F}. In the notation BEGOE(1+2)-$F$, `B' stands for bosons and $F$ is their fictitious spin-$\spin$ degree of freedom. EGOE(1+2)-$S$ has been successfully used to analyze universal conductance fluctuations in mesoscopic systems \cite{Ma-PLA}. Also, BEGOE(1+2)-$F$ may be useful in exploring general structures of spinor condensates \cite{Kaw-12}.

We analyze unitary evolution of an isolated many-body quantum system followed by an interaction quench. Initially, the system is prepared in one of the eigenstate of unperturbed mean-field Hamiltonian $h(1)$. Dynamics starts with instantaneous (in a time interval much shorter than any characteristic time scale of the model) change of interaction strength $\lambda$ from zero. Then, the final perturbed Hamiltonian is $H=h(1)+\lambda V(2)$ with eigenvalues $E$; $V(2)$ is the random perturbation. To characterize system evolution, we analyze fidelity decay and information entropy for small times. There have been several experimental \cite{Ro-06} and analytical \cite{Cam-11} studies showing that fidelity/survival probability exhibits non-exponential behavior at long times. Fidelity is related to Loschmidt echo and is defined as the overlap between the initial state and its evolved counterpart. Shannon entropy is a measure of complexity that captures participation of other ($\neq$ initial) states in system dynamics. Fidelity decay and entropy production after a quantum quench have been considered in a number of works; see for example\cite{Pe-84, FlIz, Pas-01, Go-06, Sielgman, Man-08, Be-11, Lea-12}. Going further, we analytically and numerically analyze the effect of many-particle spin on fidelity decay and entropy production. We also derive analytical formulas for saturation value and saturation time for entropy. Estimation of saturation time is important for development of algorithms for quantum optimal control \cite{Caneva}. Now we will give a preview.

Section 2 briefly introduces the EGOE(1+2)-$S$ and BEGOE(1+2)-$F$ models along with the chaos markers generated by these ensembles. Spin dependence of spectral variances is also investigated in detail as it explains the spin dependence of fidelity and entropy. Section 3 gives definitions and analytical formulas for fidelity, information entropy, saturation value and saturation time for information entropy. Numerical results are compared with analytics in Section 4. Finally, Section 5 gives conclusions. 

\section{Embedded fermionic/bosonic ensembles with spin}

To analyze time dynamics of an isolated system initially prepared in a mean-field basis state (unperturbed) followed by a random two-body interaction (non-integrable) quench, we model the system Hamiltonian by EGOE(1+2)-$S$ for fermions [and BEGOE(1+2)-$F$ for bosons]. Following our previous works \cite{Ma-PRE, Ma-F}, we briefly introduce EGOE(1+2)-$S$ [also BEGOE(1+2)-$F$] generated by one plus two-body random interactions preserving total spin $S$. 

\subsection{Definition and construction}

We consider a system of $m$ ($m >2$) fermions distributed in $\Omega$ number of single particle (sp) orbitals each with spin $\cs =\spin$ so that number of sp states is $N=2\Omega$. Distributing these $m$ fermions in $N$ sp states generates a complete set of basis states spanning the Hilbert space. As $\cs=\spin$, the two-particle spin $s=0$ or $1$ and many-particle spin $S=m/2$, $m/2-1$, $\ldots$, 0 or 1/2. EGOE(1+2)-$S$ is defined by the Hamiltonian $H$ that preserves total $m$ particle spin $S$,
\be
H= h(1) + \lambda V(2)\;;\;\;\; V(2) = V^{s=0}(2) \oplus V^{s=1}(2) \;.
\label{eq-1}
\ee
Here, $\lambda$ is the strength of the two-body interaction $V(2)$. We choose one-body Hamiltonian $h(1) = \sum_i \epsilon_i \; n_i$, specified by sp energies $\epsilon_i$ with unit average level spacing. Note that, $n_i$ are number operators acting on the sp states $i=1,\;2$, $\ldots, \;\Omega$. As two-particle spins can take two values, the two-body Hamiltonian $V(2)$ is a direct sum matrix of matrices in spin $0$ and $1$ spaces with dimensions $\Omega(\Omega+1)/2$ and $\Omega(\Omega-1)/2$ respectively. Thus, $V(2)$ is defined by two-body matrix elements that are independent of the $m_s$ quantum number. Then the $m$ particle states can be classified according to $U(N) \supset U(\Omega) \otimes SU(2)$ algebra with $SU(2)$ generating spin $S$. As $H$ preserves $S$, it is a scalar in spin $SU(2)$ space. 

EGOE(1+2)-$S$ is generated by the action of $H$ on many-particle basis states, with $V^{s=0}(2)$ and $V^{s=1}(2)$ being independent GOEs in two-particle spaces. Given the sp energies and the two-body matrix elements, the many-particle Hamiltonian $H$ is first constructed in the $M_S = M_S^{min}$ [$M_S^{min} = 0(\spin)$ for $m$ even(odd)] representation using the spin-less EGOE formalism and then states with a given $S$ are projected using a spin projection operator $S^2$ \cite{Ma-PRE, Ma-F}. Thus, the many-particle Hamiltonian $H$ will be a block diagonal matrix with each diagonal block corresponding to a given spin $S$ \cite{Ma-PRE}. Similarly, one can define BEGOE(1+2)-$F$ with $\cs$ replaced by ${\bf f}$, $s$ replaced by $f$ and $S$ replaced by $F$ \cite{Ma-F}. 

From now on, whenever we discuss the results that apply to both EGOE(1+2)-$S$ and BEGOE(1+2)-$F$, we refer to them as EE(1+2)-$S$ results.

\subsection{Transition or chaos markers}

As a function of the interaction strength $\lambda$, the embedded ensembles with one plus two-body interactions exhibit three transition or chaos markers $(\lambda_c,\lambda_\ell,\lambda_d)$: (a) as the two-body interaction is turned on, level fluctuations exhibit a transition from Poisson to GOE at $\lambda=\lambda_c$; (b) with further increase in $\lambda$, the strength functions (also known as the local density of states) make a transition from Breit-Wigner (BW) to Gaussian form at $\lambda=\lambda_\ell > \lambda_c$; and (c) beyond $\lambda=\lambda_\ell$, there is a region around $\lambda=\lambda_d$ where basis dependent thermodynamic quantities like entropy behave alike. For further details of the three chaos markers for EE(1+2) for spinless systems, see \cite{Ko-14, KoReta, Ch-PLA}. It was established that EE(1+2)-$S$ also exhibits these chaos markers and their spin dependence was also deduced \cite{Ma-PRE, Ma-F}. Exact formula for the ensemble averaged variances [second moment of the eigenvalue density with fixed-$(\Omega,m,S)$] was also derived using trace propagation method. Importantly, the variances explain the spin dependence of various chaos markers. 

\subsection{Spin dependence of spectral variances}

Initially, the system is in eigenstate $\Psi(t=0) = \l.\l| k\r.
\ran$ of unperturbed mean-field Hamiltonian $h(1)$. The dynamics begin with the 
sudden change in the parameter $\lambda$, the strength of the perturbation, in a time interval much shorter than any characteristic time scale of the model.
With the quench $V(2)$ of strength $\lambda$, the state unitarily changes after time $t$ to $\Psi(t)$ given by
\be
\Psi(t) = \l.\l| k(t) \r.\ran = \exp -iHt\,\l.\l| k\r.\ran\;.
\label{eq.time3}
\ee
Here, $H$ is given by Eq. (\ref{eq-1}). Then, the probability that the state $\l.\l| k \r.\ran$ changes to the state $\l.\l| f \r.\ran$ is $W_{k \rightarrow f}(t)$,
\be
\barr{rcl}
W_{k \rightarrow f}(t) & = & \l| \lan f \mid k(t) \ran \r|^2 =
\l| A_{k \rightarrow f}(t) \r|^2 \;;\\ \\
A_{k \rightarrow f}(t) & = & \dis\sum_E \l(C_f^E\r)^* C_k^E  \exp -iEt \;,
\earr \label{eq.time5}
\ee
where the coefficients $C_{--}^E$ are the overlaps of mean-field basis states ($\l.\l| k \r.\ran$ and $\l.\l| f \r.\ran$) with the eigenstates $\l| E \ran$ of the post-quench Hamiltonian $H$.  We use $\hbar=1$ so that $t$ is in $E^{-1}$ units. It is important to mention that, although not explicitly included in the notation, the energy eigenstates $\l| E \ran$ of $H$ depend on many-particle spin $S$($F$) and thus, $W_{k \to f}(t)$ and $A_{k \to f}(t)$ too.

As we will see ahead in Section 3, the ensemble averaged spectral variance $\overline{\sigma^2_k(m,S)}$ of the initial basis states $|k\rangle$ is the
single most crucial parameter in the EGOE(1+2)-$S$ [also BEGOE(1+2)-$F$] formalism to analytically understand the spin dependence of fidelity and entropy.  
Firstly as argued in \cite{Ks-01}, $\overline{\sigma^2_k(m,S)}$ will be essentially independent of $k$ and therefore for EGOE(1+2)-$S$ for fermionic systems, we have
\be
\overline{\sigma^2_k(m,S)}=\lambda^2\;\frac{\overline{\sigma^2_{V(2)}(m,S)}}{
\overline{\sigma^2_H(m,S)}}\;\;.
\label{eq.sigmak}
\ee
Note that `overline' denotes ensemble average and $\sigma^2_{V(2)}(m,S)$ and $\sigma^2_H(m,S)$ are variances for the perturbation $V(2)$ and the Hamiltonian $H$ respectively. Assuming that $h(1)$ is fixed, the ensemble averaged spectral variance $\overline{\sigma^2_H(m,S)}$ follows simply from Eq. \eqref{eq-1}, 
\be
\overline{\sigma^2_H(m,S)}= \sigma^2_{h(1)}(m,S) +\lambda^2\; 
\overline{\sigma^2_{V(2)}(m,S)}\; \;.
\label{eq.sigmaH}
\ee
Now, for a uniform sp spectrum having unit average level spacing, the 
$\sigma^2_{h(1)}(m,S)$ follows easily from Eq. (7) of \cite{Ma-PRE},
\be
\sigma^2_{h(1)}(m,S) = \frac{1}{12}\Big[m(\Omega+2)(\Omega-m/2)-
2\Omega S(S+1) \Big]\;.
\label{eq.sigmah}
\ee
Similarly, the ensemble averaged variance generated by the perturbation $V(2)$
is $\overline{\sigma^2_{V(2)}(m,S)} = P(\Omega, m, S)$. For BEGOE(1+2)-$F$, Eq. (\ref{eq.sigmaH}) with $S$ replaced by $F$ applies and the variance propagator is $Q(\Omega, m, F)$. Explicit formulas, derived using trace propagation method, for the variance propagators $P(\Omega, m, S)$ and $Q(\Omega, m, F)$ are given in \cite{Ma-PRE} and \cite{Ma-F} respectively. For sake of completeness, we give the exact formulas for these in the Appendix. Again, for a uniform sp spectrum with unit mean level spacing, $\sigma^2_{h(1)}(m,F)$ follows from Eq. (\ref{eq.sigmah}) by $\Omega \to -\Omega$ transformation and applying modulo operation,
\be
\sigma^2_{h(1)}(m,F) = \frac{1}{12}\Big[m(\Omega-2)(\Omega+m/2) +
2\Omega F(F+1) \Big]\;.
\ee
For simplicity of notation, from now on, we drop the `bar' over $\overline{\sigma^2_k}$ as we will only deal with ensemble averages. Note that variances are always positive as for given $(\Omega,m)$, the many-particle spin can only take the allowed values $m/2$, $m/2-1$, $\ldots$, 0 or $\spin$.

\begin{figure*}[!ht]
\begin{center}
\begin{tabular}{cc}
\includegraphics[width=0.5\linewidth]{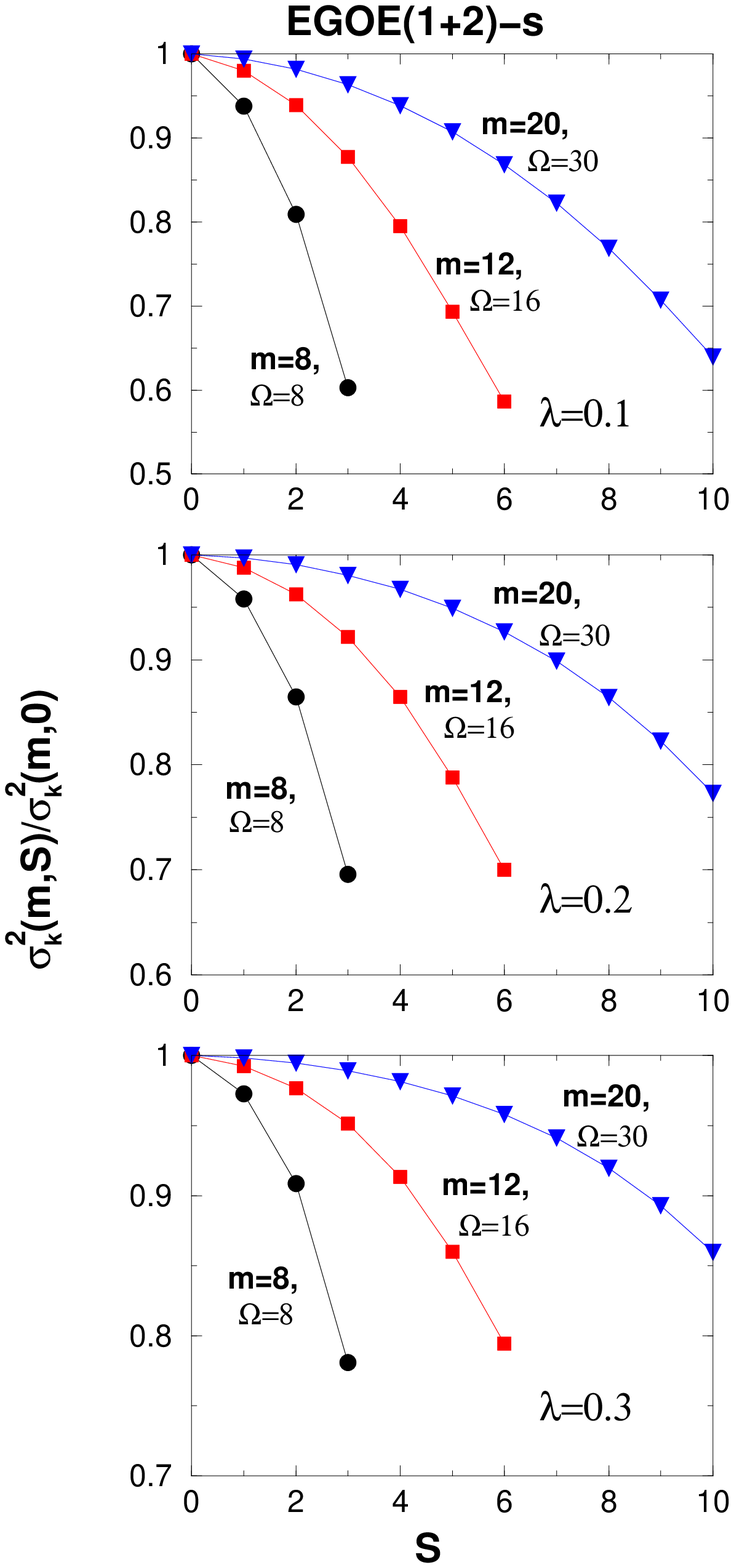} & 
\includegraphics[width=0.5\linewidth]{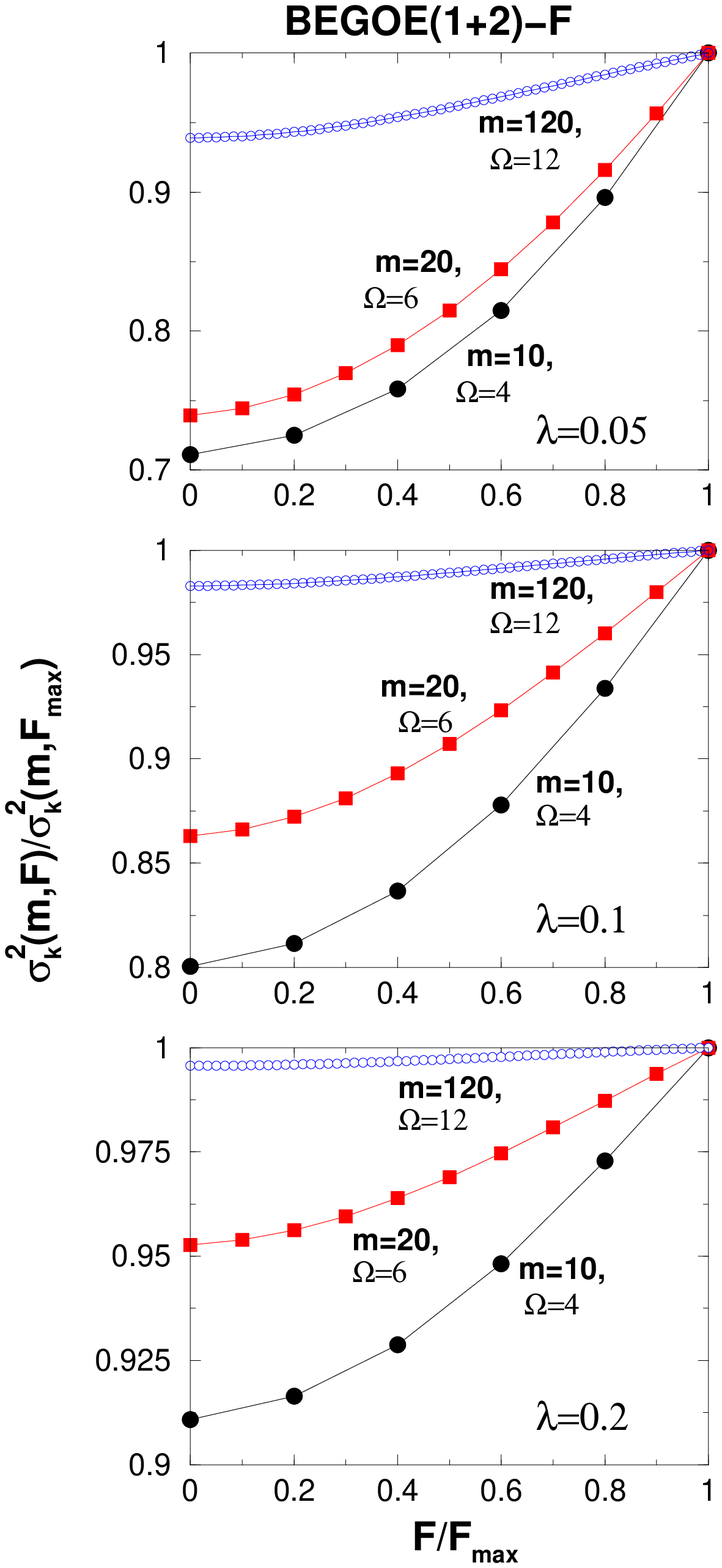}  \\
(a) & (b)
\end{tabular}
\end{center}
\caption{Variation in the ensemble averaged spectral variances of the initial states $\sigma^2_k(m,S)$ for various $(\Omega, m, \lambda)$ values for (a) EGOE(1+2)-$S$ and (b) BEGOE(1+2)-$F$ as a function of many-particle spin $S$($F$). See text for further details.}
\label{fig.sigma}
\end{figure*}

General behavior of $\sigma^2_k(m,S)$ with $S$ for fermionic systems, using Eqs. (\ref{eq.sigmak})-(\ref{eq.sigmah}) and the formula for variance propagator $P(\Omega,m,S)$ given by Eq. (\ref{appndx-1}), is shown for some values of $(\Omega,m, \lambda)$ in Fig.~\ref{fig.sigma}a. It is seen that $\sigma^2_k(m,S)$ decreases with increasing spin as the number of states connected with each other decrease. Similarly, the results for BEGOE(1+2)-$F$ are shown in Fig.~\ref{fig.sigma}b. As seen from the figure, the trend is opposite for bosonic systems. 

In order to understand the scaling of $\sigma^2_k(m,S)$ with system size $m$, we examine the asymptotic form of $\sigma^2_k(m,S)$. For EGOE(1+2)-$S$, for fixed density $\rho_0 ={m}/{\Omega}$ with $m \to \infty$ and $S=m/2$ (the maximum value) we have,
\be
\sigma^2_k\l(m,\frac{m}{2}\r) = 1-\dis\frac{1}{3 \lambda^2 (1-\rho_0) m} + O\l(\frac{1}{m^2}\r)\;.
\ee
Similarly for minimum spin $S=0$ (assuming $m$ is even) we have,
\be
\sigma^2_k(m,0) = 1-\dis\frac{2}{3 \lambda^2 (2-\rho_0) m} + O\l(\frac{1}{m^2}\r) \;.
\ee
For $S$ varying from $0$ to $m/2$ for $m$ even ($\spin$ to $m/2$ for $m$ odd), the $\sigma^2_k(m,S)$ varies between the above two limits. It should be noted that for an arbitrary and fixed $S$, the expansions are complicated as one has to include $O(1/m^p)$ terms with $p$ increasing with $S$.

This behavior is reflected in the fidelity decay and hence in longer time for relaxation. These are discussed further in the next Section.

\section{Fidelity and entropy: Definitions and EE(1+2)-$S$ formulas}

Both fidelity and information entropy are basis dependent quantities. In our construction of the $H$ matrices, the chosen basis states are eigenstates of both $h(1)$ and $S^2$ operators (although all the basis states have $M_S=M_S^{min}$, we drop $M_S^{min}$ everywhere).

\subsection{Fidelity}

Probability to find the system still in the initial state $\l| k \ran$ after time $t$ corresponds to fidelity (also referred to as the return or the survival probability).
It is given by,
\be
\barr{rcl}
W_{k \rightarrow k}(t) & = & \l| A_{k \rightarrow k}(t) \r|^2 =
\l| \dis\sum_E \l[C_k^E\r]^2
\exp -iEt \r|^2 \\ \\
& = & \l| \dis\int F_k^{m,S}(E) \exp-iEt\;dE \r|^2 .
\earr
\label{eq.time6}
\ee
Here $F_k^{m,S}(E)=|C_k^E|^2\,\rho^{m,S}(E)$ are the strength functions or the local density of states and $\rho^{m,S}(E)$ is the normalized eigenvalue density with fixed-$(\Omega,m,S)$. Strength functions give the spreading of the basis states over the eigenstates and thus, fidelity is the Fourier transform in energy of the strength function $F_k^{m,S}(E)$. Our focus is on the general scenario with single peaked energy distribution of the initial states. 

Following our discussion in Sec. 2.2, we know that EE(1+2)-$S$ exhibits three chaos markers $(\lambda_c,\lambda_\ell,\lambda_d)$. As a function of the perturbation strength $\lambda$, the strength functions are $\delta$-functions at the $h(1)$ eigenvalues for very small $\lambda$. With increasing $\lambda$, the eigenstates start spreading over all the basis states leading to delocalization in the Fock basis. First, they become BW in shape and finally, acquire Gaussian form around and beyond $\lambda = \lambda_\ell$ \cite{Ma-PRE}. For $\lambda > \lambda_c$, the level and strength fluctuations follow GOE and hence, in this region one can replace, to a good approximation, $F_k(E)$ by its smoothed (ensemble averaged) form. As the smoothed form of $F_k(E)$ changes from BW to Gaussian with increasing $\lambda$, there are four situations: (i) small `$t$' limit where one can apply perturbation theory; (ii) BW region;  (iii) Gaussian region; (iv) region intermediate to BW and Gaussian forms. Results for (i)-(iii) are already discussed by Flambaum and  Izrailev~\cite{Fl-Aust,FlIz} and briefly they are as follows.

For very small times, we have $\exp -iHt \simeq \l[\exp -ih(1)t\r]\, \l[\exp
-i \lambda V(2)t\r]$. Using $e_k(m,S) = \lan k \mid H \mid k\ran \simeq \lan k \mid h(1) \mid k \ran$, $\sigma^2_k(m,S) = \lan k \mid H^2 \mid k \ran -
e_k^2(m,S) \simeq \lan k \mid [\lambda V(2)]^2 \mid k \ran$ and the dominant terms (up to second order) in the expansion in time of $\exp[-i \lambda V(2)t]$, the fidelity is, $W_{k \rightarrow k}(t) = 1 - \sigma_k^2(m,S)\; t^2$. 
The long time behavior of fidelity decay depends also on the shape of the distribution of $F_k^{m,S}(E)$. For $\lambda$ not far from $\lambda_c$, the $F_k^{m,S}(E)$ has BW distribution with level and strength fluctuations following GOE. Thus, fidelity, as seen from Eq. (\ref{eq.time6}), follows exponential law: $W_{k \rightarrow k}(t)  \longrightarrow \exp -\Gamma(m,S) t$. Here, $\Gamma(m,S)$ is the spreading width of the BW distribution. Similarly, in the Gaussian region, we obtain $A_{k \rightarrow k}(t) =\exp -\l[i e_k(m,S)\; t +\sigma_k^2(m,S) \; t^2/2\r]$ using Eq. (\ref{eq.time6}). Therefore, fidelity follows Gaussian law: $W_{k \rightarrow k}(t) \longrightarrow \exp -\sigma_k^2(m,S) \; t^2$. Note that when $t$ is measured in $[\sigma_H(m,S)]^{-1}$ units, the spreading widths $\Gamma(m,S)$ and $\sigma_k(m,S)$, respectively of the BW and Gaussian distributions, will be in units of $\sigma_H(m,S)$. In summary,
\be
\barr{rcl}
W_{k \rightarrow k}(t) & \stackrel{\mbox{BW region}}{\longrightarrow} & \exp
-\Gamma(m,S) \, t\;, \\ \\
W_{k \rightarrow k}(t) & \stackrel{\mbox{Gaussian region}}{\longrightarrow} & \exp
-\sigma_k^2(m,S) \, t^2 \;.
\earr \label{eq.time15}
\ee
Therefore, the logarithm of fidelity decay $(\ln W_{k \to k}(t))$ shows a linear dependence on time in the BW region while the dependence is quadratic in the Gaussian regime.

In the BW to Gaussian transition region, as demonstrated by Angom et
al~\cite{An-04},  it is possible to approximate $F_k(E)$ by Student's ${\bf t}$-distribution in terms of the shape parameter $\alpha$ and scale parameter
$\beta$. With the transformations $\alpha=(\nu +1)/2$ and
$[E-e_k(m,S)]=\sqrt{\frac{\beta (\nu +1)}{2\nu}}\,x$, the $F_k^{m,S}(E)$ in the BW to Gaussian transition region transforms to $F_k^{m,S}(x:\nu)$ where,
\be
F_k^{m,S}(x:\nu)=\dis\frac{\Gamma\l(\frac{\nu +1}{2}\r)}{\dis\sqrt{\pi}\,
\dis\sqrt{\nu}\,\Gamma\l(\frac{\nu}{2}\r)}\;{\l(\dis\frac{x^2}{\nu} +1\r)^{-\dis\frac{(\nu +1)}{2}}} \; {dx}\;.
\label{eq.time16}
\ee
To analytically compute the fidelity in  the BW to Gaussian transition region, 
we need the Fourier transform of $F_k^{m,S}(x:\nu)$. This is a topic of interest in many statistics problems. Using the result given in~\cite{DK-02}, we get,
\begin{widetext}
\be
\barr{l}
W_{k \rightarrow k}(t) \stackrel{\mbox{transition region}}{\longrightarrow}
\l| \dis\frac{2^\nu\,
\l(\dis\sqrt{\nu}\r)^\nu}{\Gamma(\nu)}\;\dis\int_{0}^{\infty}dx
\l[x(x+\l| t^\prime \r|)\r]^{(\nu -1)/2}\; e^{-\dis\sqrt{\nu}\,(2x+
\l| t^\prime \r|)}\,\r|^2\;;\\ \\
t^\prime = \dis\sqrt{\dis\frac{\beta (\nu +1)}{2\nu}}\,t\;.
\earr 
\label{eq.time18}
\ee
\end{widetext}
We correctly recover the results for BW and Gaussian limits using Eq. (\ref{eq.time18}) as well: (a) for $\nu=1$, we get BW form for $F_k^{m,S}(E)$ with $\beta=\Gamma^2/4$ and (b) for $\nu \to \infty$, we get Gaussian form with $\sigma^2_k(m,S)=\beta/2$. In general, $\sigma_k^2(m,S)$ is related to the parameters $(\alpha,\beta)$ by  
$\sigma^2_k(m,S)=\frac{\alpha}{2\alpha-3}\beta; \; \alpha > 3/2$. Eq. (\ref{eq.time18}) was first used in \cite{Ko-14} to analyze fidelity decay in the framework of EE(1+2). Thus, we have full EE(1+2)-$S$ theory for
fidelity decay for different regimes defined by perturbation strength $\lambda$ in terms of $S$ dependent spectral widths.

\subsection{Entropy}

Information entropy $S(t)$ (also known as Shannon entropy) is a measure of complexity or disorder. Assuming that there are total $D+1$ mean-field basis states such that $f=0,1, \ldots,D$ with $\l.\l|f=0\r.\ran=\l.\l|k\r.\ran$, the initial state in which the system is prepared at time $t=0$, information entropy after time $t$ is
\begin{equation}
S(t) = - \dis\sum_{f=0}^D\,W_{k \to f} (t) \ln W_{k \to f}(t)\;.
\label{eq.time19}
\end{equation}
It is important to recognize that
\begin{equation}
\dis\sum_{f=0}^D\;W_{k \to f}(t) =1\;.
\label{eq.time20}
\end{equation}
Using Eq. (\ref{eq.time5}), we have
\begin{equation}
\begin{array}{rcl}
W_{k \to f}(t) & = & \dis\sum_{E} \l|C_0^E \r|^2 \,\l|C_f^E \r|^2 
 \\ \\
& + & 
2 \dis\sum_{E > E^\pr}
C_0^E C_f^E C_0^{E^\pr} C_f^{E^\pr}\;\cos(E-E^\pr) \; t \\ \\
& = & W_{k \to f}^{avg}(t) + W_{k \to f}^{flu}(t) \;.
\end{array}
\label{eq.time21}
\end{equation}
As system at time $t=0$ is prepared in initial state $\l.\l|k\r.\ran$, for simplicity of notation, henceforth we denote $W_{k \to k}(t) = W_0(t)$. Note that formulas for $W_0(t)$ are derived in Sec. 3.1 fully taking into account both the terms in Eq. (\ref{eq.time21}). 

Isolated interacting many-body quantum systems equilibrate in a probabilistic way. After a long time ($t \to \infty$), the local observables fluctuate around their infinite time-averages with the size of the fluctuations decreasing with increasing system size. Therefore, it is plausible to argue that the second term in Eq. (\ref{eq.time21}) that gives fluctuations for $f \neq 0$ approaches zero for very long times. Similarly, in the short time limit ($t$ close to zero),  $S(t) \longrightarrow \sigma_k^2(m,S) t^2 - t^2 \sum_{f=1}^D H^2_{0f} \ln \l\{H_{0f}^2 t^2\r\}$, i.e. entropy $S(t)$ will be quadratic in $t$. Here, $H^2_{0f} = |\lan f \mid H \mid 0\ran|^2$ are like transition strengths. For $t$ neither very small and very large, EE(1+2) theory for $S(t)$ is not available. Earlier, using a cascade model as an approximation, Flambaum and Izrailev~\cite{FlIz} showed that entropy initially increases linearly with $t$ and eventually saturates. In this model, Given a set of states, these are separated into different classes with states directly coupled with those from the previous class populating the successive class. However, one needs to extend the Cascade model to account for low connectivities as well as different connectivities for different classes. 

To derive approximate EE(1+2)-$S$ formulas for $S(t)$, its saturation value and the time $t_{sat}$ which marks the onset of its saturation, assume that in the sum in Eq. (\ref{eq.time19}), variation of $W_{k \to f}(t)$ with $f$ [i.e. the fluctuations in $W_{k \to f}(t)$] is negligible. This is physical as for sufficiently large interaction strength $\lambda > \lambda_c$, the system becomes chaotic and there is no preference to any particular mean-field basis state. Then, one can replace $W_{k \to f}$ by its average value $\overline{W}$. Suppose that there are $n$ number of mean-field basis states $f$ with $f \neq 0$ which contribute to the sum in Eq. (\ref{eq.time19}). Applying Eq.~(\ref{eq.time20}) using the identity $\sum_{f\ne 0} 1 = n$, we have
\be
\barr{l}
S(t) = -W_0(t) \ln W_0(t) - \dis\sum_{r=1}^{n} \overline{W} \ln \overline{W} 
\;;\;\;\;\overline{W}=\dis\frac{1-W_0}{n} \\
\\
= -W_0(t) \ln W_0(t) - \l[1-W_0(t)\r] \ln \l(\dis\frac{1-W_0(t)}{n}\r)\;. 
\earr 
\label{eq.time30}
\ee
Now, we will derive the EE(1+2)-$S$ formulas for $t_{sat}$ in Gaussian, BW and BW to Gaussian transition regimes using Eq. (\ref{eq.time30}).

\subsubsection{Gaussian regime}

In the Gaussian regime $\lambda > \lambda_\ell$, fidelity $W_0(t)$ follows Gaussian law; see Eq.~(\ref{eq.time15}). Substituting this in Eq.~(\ref{eq.time30}), we have
\begin{widetext}
\be
S(t) = \sigma_k^2(m,S) \, t^2 \, e^{-\sigma_k^2(m,S) t^2}
- \l[1-e^{-\sigma_k^2(m,S) t^2}\r] \ln\l(\dis\frac{1-e^{-\sigma_k^2(m,S) t^2}}{n}\r)\;.
\label{eq.theory}
\ee
\end{widetext}
It is quite clear that $n=e^{S(\infty)}$, with $S(\infty)$ being the long time saturation value of entropy.  Ideally one needs to derive a formula for $S(\infty)$ by taking the limit $t \rightarrow \infty$ in Eq.~(\ref{eq.time19}). However, this is not solved yet. it is plausible to expect that $n$ should be  
proportional to the number of mean-field basis states over which the eigenstate spreads. This is generally referred to as number of principal components (NPC). NPC is expected to be large for chaotic systems (Gaussian regime) and hence, these systems will thermalize unlike regular ones. The maximal value of NPC for EE(1+2)-$S$ in Gaussian regime is given by putting $\hat{E}=0$ in Eq. (5) of \cite{Ks-01}. Therefore, approximating $n$ by the maximal value of NPC for EE(1+2)-$S$,
\begin{equation}
\barr{l}
n \sim \kappa (\rm{NPC}_{max}) = \kappa \dis\frac{d(m,S)}{3}\dis\sqrt{1-\zeta^4(m,S)}\;; \\ \\
\zeta^2(m,S) = 1- \sigma^2_{k}(m,S) \;.
\earr
\label{eq.ns-th}
\end{equation}
Here, $\zeta^2(m,S)$ is a correlation coefficient and $\kappa$ is a parameter.  
Eqs.~(\ref{eq.theory}) and (\ref{eq.ns-th}) give a theoretical description of time evolution of the entropy for EE(1+2)-$S$ in the Gaussian regime. For very short times, expanding the exponential and retaining only dominant contributions in Eq. (\ref{eq.theory}), entropy shows a quadratic growth in time.

\begin{figure*}[htp]
\begin{center}
\includegraphics[width=\linewidth]{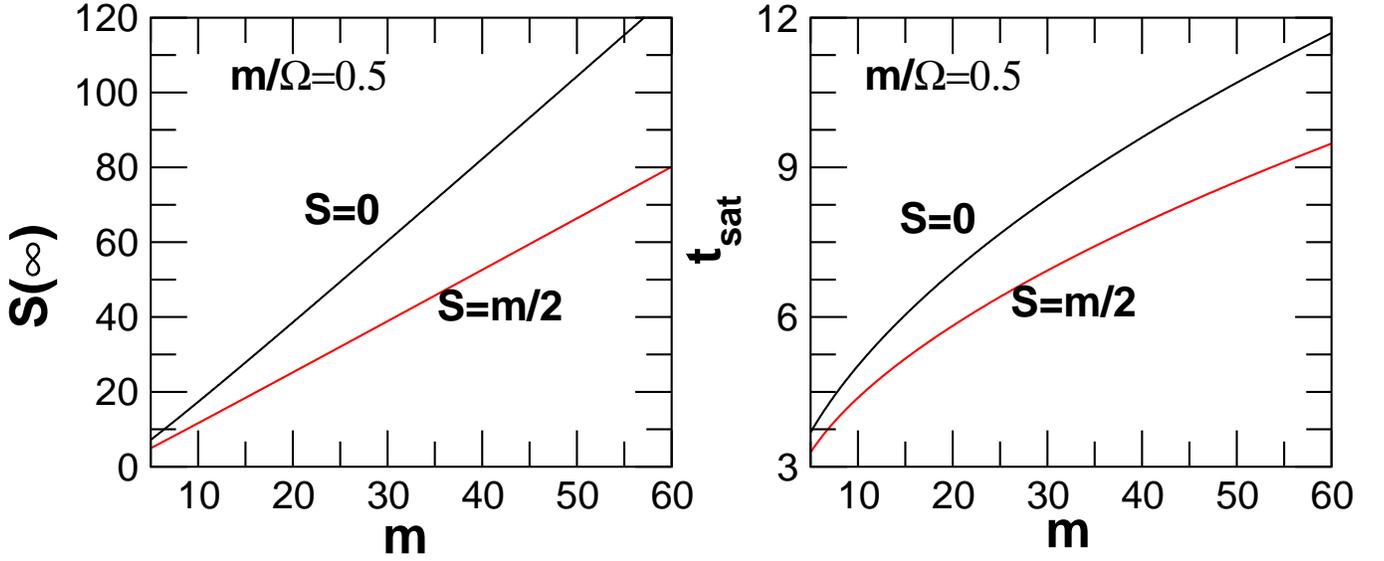} 
\end{center}
\caption{Asymptotic behavior of $S(\infty)$ (left panel) and $t_{sat}$ (right panel) with system size $m$. Results are shown for EGOE(1+2)-$S$ ensemble with $\rho_0 = m/\Omega =0.5$,  $\lambda=0.3$ and many-particle spins $S=0$ and $m/2$.}
\label{fig.tsat-sinfty}
\end{figure*}

We define the saturation time as the time taken by the information entropy to reach its saturation value. Upon saturation, the time variation of entropy vanishes. Thus, setting time derivative of $S(t)$ given by Eq.~(\ref{eq.theory}) to zero, we obtain
\begin{widetext}
\begin{equation}
\begin{array}{l}
2 \, \l[1-\zeta^2(m,S)\r] \, t^2 \, e^{-[1-\zeta^2(m,S) ]t^2} \l\{\l[1-\zeta^2(m,S)\r] \, t^2 + 
\ln\Big(\dis\frac{1-e^{-\l[1-\zeta^2(m,S)\r] t^2]}}{n}\Big) \r\} = 0 \;.
\label{eq.eqty}
\end{array}
\end{equation}
\end{widetext}
Simplifying Eq. (\ref{eq.eqty}), we get
\begin{equation}
 t_{sat} = \sqrt{\frac{\ln(1+n)}{1-\zeta^2(m,S)}}\;.
\label{eq.t-sat}
\end{equation}
As $\lambda \rightarrow \infty$, $\zeta^2(m,S)$ approaches $0$ and we get a
lower bound for $t_{sat}$, 
\be
t_{sat}^{min} \simeq \dis\sqrt{\ln \Big(\kappa\;\dis\frac{d(m,S)}{3}\Big)}\;.
\label{asymp}
\ee
Importantly, $\zeta^2$ approaches $1$ for $\lambda \rightarrow 0$ giving
$t_{sat} \rightarrow \infty$. Thus, within the EE(1+2)-$S$ formulation given here, an integrable system will never thermalize. As discussed in Sec. 2.2, an isolated closed quantum system thermalizes around the transition marker $\lambda =\lambda_d$. In this case, the spreadings produced by the mean-field basis and the interaction basis are equal and hence, $\zeta^2(m,S)=0.5$ that gives the upper bound for the time a chaotic system takes to thermalize,  $t_{th} \simeq \sqrt{2 \ln(\kappa  \frac{d(m,S)}{6}\sqrt{3}})$.

\subsubsection{BW regime}

In the BW region $\lambda_c < \lambda < \lambda_\ell$, fidelity decay $W_0(t)$ is exponential in time; see Eq. (\ref{eq.time15}). Using this in Eq.~(\ref{eq.time30}) and proceeding in the same way as in Sec. 3.2.1, we obtain
\be
t_{sat} = \dis\frac{\ln(1+n)}{\Gamma(m,S)}\;.
\label{eq.t-sat-bw} 
\ee

\subsubsection{BW to Gaussian transition region}

In the BW to Gaussian transition region, $W_0(t)$ is given by Eq.~(\ref{eq.time18}) and the formula for $\rm{NPC}_{max}$ was derived in~\cite{An-04}. Again, approximating $n$ by $\rm{NPC}_{max}$, we have
\begin{widetext}
\be
\barr{rcl}
n & \sim & \kappa (\rm{NPC}_{max}) \\ \\
& = & \kappa \dis\frac{d(m,S)}{3}\l\{\dis\sqrt{\dis\frac{2}{2\alpha-3}}
\dis\frac{\Gamma^2(\alpha)}{\Gamma^2(\alpha-\dis\frac{1}{2})}\dis\frac{1}{
\dis\sqrt{\zeta^2(m,S)[1-\zeta^2(m,S)]}}
U\l(\frac{1}{2},\frac{3}{2}-2\alpha,\frac{(2\alpha-3)[1-\zeta^2(m,S)]}{2\zeta^2(m,S)}
\r)\r\}^{-1}\,.
\earr
\label{eq.npc-int}
\ee
\end{widetext}
In Eq. (\ref{eq.npc-int}), $\rm{U}(---)$ is the hypergeometric-$U$ function~\cite{Abramowitz} and $\Gamma(-)$ are the Gamma functions. 
Time derivative of entropy shows that $t_{sat}$ should be
inversely proportional to  $\l(\frac{\beta}{2}\r)^x\l(1+\frac{1}{\nu}\r)^y$.
Also, $x$ and $y$ should be chosen such that $t_{sat}$ converges to Eqs.~(\ref{eq.t-sat}) and (\ref{eq.t-sat-bw}) in the $\nu \rightarrow \infty$ 
and $\nu=1$ limits respectively. Using these, $t_{sat}$ in the BW to Gaussian transition region is
\be
t_{sat}=\dis\frac{\l[\ln \l(1+n\r)\r]^{\dis\frac{1}{2}\l(1+\dis\frac{1}{\nu}\r)}}{\dis\sqrt{
\dis\frac{\beta}{2}}\Big(1+\dis\frac{1}{\nu}\Big)^{3/2}}\;; \nu=2\alpha -1 \;.
\label{eq.tsat-int}
\ee

For EGOE(1+2)-$S$, we have also derived tedious analytical expressions for $S(\infty)$ and $t_{sat}$ in the asymptotic limit with fixed density $\rho_0 = m/\Omega$ with $m \to \infty$. Fig.~\ref{fig.tsat-sinfty} shows the asymptotic variation of saturation values $S(\infty)$ (left panel) and saturation times $t_{sat}$ (right panel) of information entropy with system size $m$. With increasing system size, both $S(\infty)$ and $t_{sat}$ increase. 

Earlier, equilibration of isolated quantum system has been studied in the context of spin-$1/2$ lattice models conserving $S_z$, total spin in the $z$ direction~\cite{Manan}. A relaxation time $t_R$ has been defined as the time required by fidelity to reach its long time average value. For a two-body interaction and initial state energy close to the middle of the spectrum, a lower limit for the $t_R$ was also estimated in~\cite{Manan} and this turns out to be very close to $t_{sat}$ given by Eq.~(\ref{eq.t-sat}). However, it should be clear that $t_{sat}$ is not a bound on the saturation time and it is exact within the EE(1+2)-$S$ formalism. Rather it turns out be a relatively higher estimation of the saturation time as can be seen from numerical results ahead. 

\section{Numerical results}

\begin{figure*}[!ht]
\begin{center}
\includegraphics[width=0.9\linewidth]{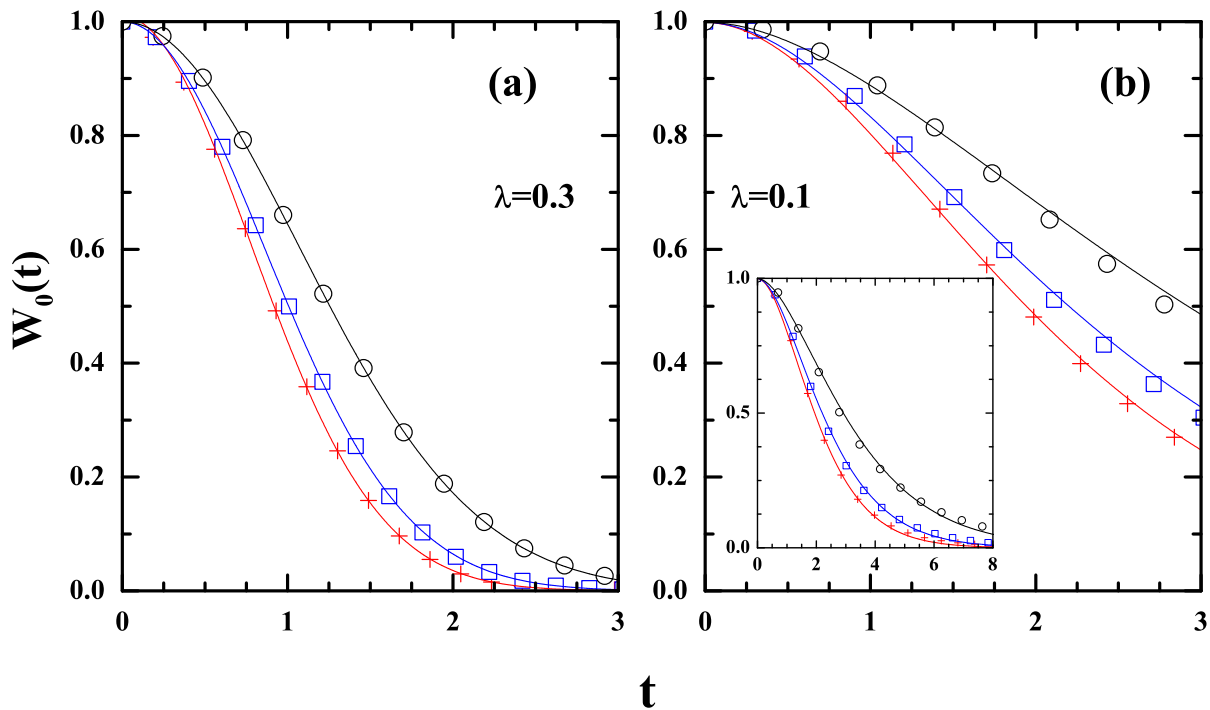} 
\end{center}
\caption{Fidelity decay $W_0(t)$ with time as a function of many-particle spins $S$ for an  EGOE(1+2)-$S$ ensemble in the: (a) Gaussian region ($\lambda=0.3$) and (b) BW to Gaussian transition region ($\lambda=0.1$). The points marked by red '+' ($S=0$),  blue 'open square' ($S=1$) and black 'open circle' ($S=2$) correspond to the numerical results. The solid continuous curves  represent the corresponding theoretical results for different $S$ values given by  Eq.~(\ref{eq.time15}) and Eq.~(\ref{eq.time18}) for panels (a) and (b) respectively.  Inset to panel (b) gives long time variation (up to $t=8$) in fidelity decay in the BW to Gaussian regime.}
\label{fig.w0-f}
\end{figure*}

\begin{figure*}[!ht]
\begin{center}
\includegraphics[width=0.9\linewidth]{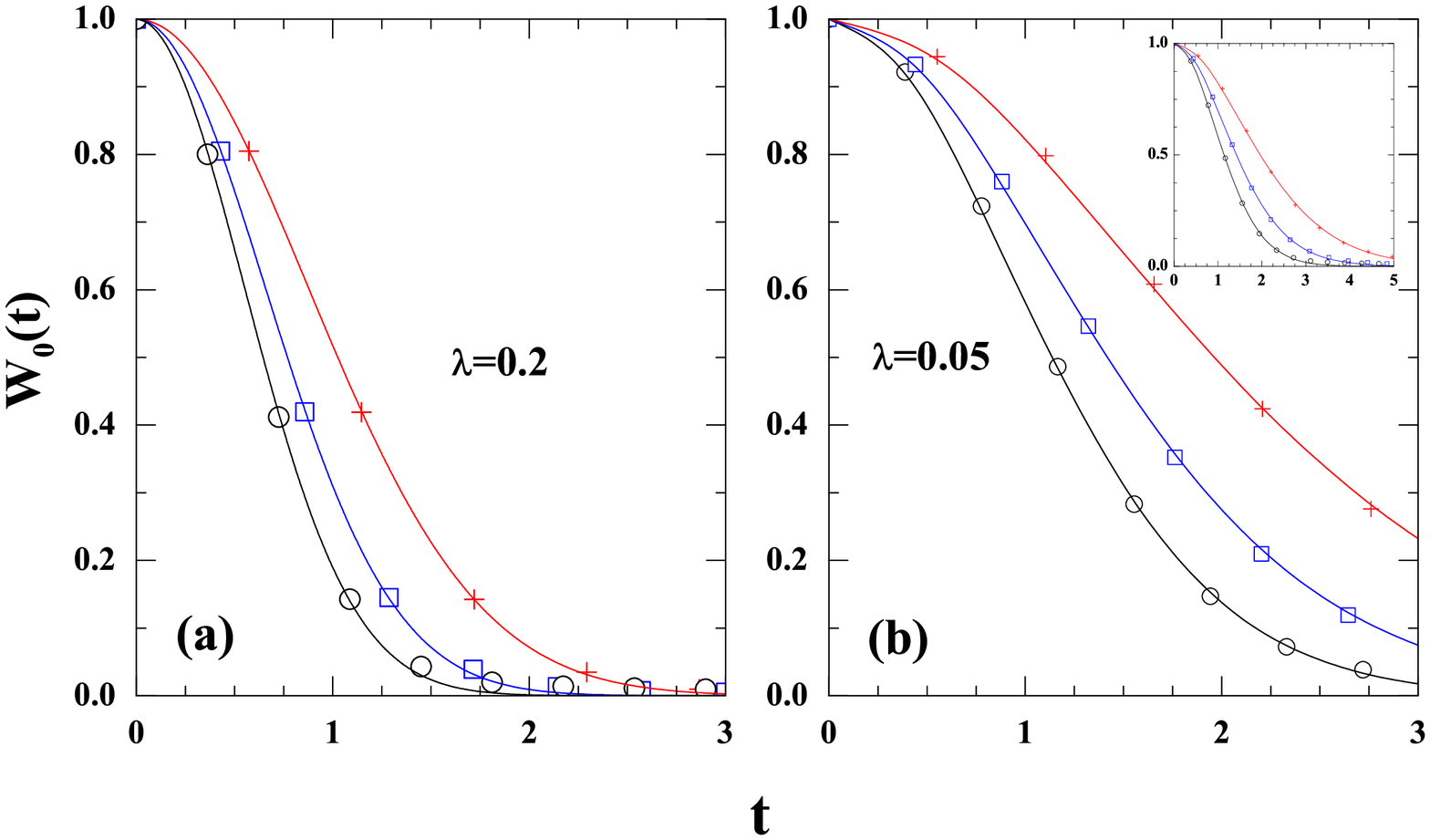} 
\end{center}
\caption{Fidelity decay $W_0(t)$ with time as a function of many-particle spins $F$ for a BEGOE(1+2)-$F$ ensemble in the: (a) Gaussian region ($\lambda=0.2$) and (b) BW to Gaussian transition region ($\lambda=0.05$). The points marked by red '+' ($F=2$),  blue 'open square' ($F=4$) and black 'open circle' ($F=5$) correspond to the numerical results. The solid continuous curves  represent the corresponding theoretical results for different $F$ values given by  Eq.~(\ref{eq.time15}) and Eq.~(\ref{eq.time18}) for panels (a) and (b) respectively. Inset to panel (b) gives long time variation (up to $t=5$) in fidelity decay in the BW to Gaussian regime.}
\label{fig.w0-b}
\end{figure*}

For the numerical calculations reported for fidelity and entropy in this Section, we make use of the following parameters. We choose fixed sp energies $\epsilon_i=i+1/i$ as in our previous studies \cite{Ma-PRE, Ma-F} and generate 20 member ensembles. For EGOE(1+2)-$S$ ensemble with $\Omega=m=8$ and spins $S=0-2$, dimensions are $d(m,S) = 1764$, $2352$ and $720$. Note that $\sum_{S=S_{min}}^{S_{max}} (2S+1) \; d(m,S) = {{N}\choose{m}}$. For this choice of parameters, (a)  $\lambda_c= 0.028$, $0.034$ and $0.05$; (b) $\lambda_\ell = 0.15$, $0.16$ and $0.19$; and (c) $\lambda_d= 0.21$, $0.22$ and $0.24$; for $S=0-2$ \cite{Ma-PRE}. Similarly, for BEGOE(1+2)-$F$ ensemble with $\Omega=4$, $m=10$, dimensions are $d(m,F) = 750$, $594$ and $286$ for $F=2$, 4 and 5 respectively; $\sum_{F=F_{min}}^{F_{max}} (2F+1) \; d(m,F) = {{N+m-1}\choose{m}}$. Here, the $\lambda_c$ values  are $0.039$, $0.0315$ and $0.0275$ for $F=0$, $2$ and $5$ respectively \cite{Ma-F}.  

First of all, the basis state energies $e_k(m,S)$ are the diagonal elements of the $H$ matrix in the many-particle Fock-space basis giving $e_k(m,S)= \lan k \l|h(1)+ \lambda V(2) \r|k\ran$. Note that the centroids of the $e_k(m,S)$ energies
are the same as that of the eigenvalue ($E$) spectra but their widths are
different. For each member of the ensemble, energies $E$ and $e_k(m,S)$ are zero centered and scaled by the spectrum width $\sigma_H(m,S)$. Then, for each member at a given time $t$, $|A_{k \rightarrow f}|^2$ are summed over the basis states $\l|k\ran$ and $\l|f \ran$ in the energy windows $e_k \pm \delta_1$ and $e_f \pm \delta_2$. Here, $\delta_1$ and $\delta_2$ are the bin-sizes and they are chosen such that there are approximately 4-7 states within each bin. In principle, they can be different or same depending on the choice of initial and final Hilbert spaces. We choose $\delta_1 =\delta_2 = 0.01$ and $e_k(m,S)=0$. Numerical calculations, though not reported, were also performed with $\delta_1 = \delta_2 = 0.05$ and $0.1$ and it was confirmed that the results remain unaffected. Then, ensemble averaged transition probability $W_{k \rightarrow f}$ for a fixed initial mean-field basis state $k$ is obtained by binning. Note that this procedure is different from the one adopted in \cite{FlIz} where no ensemble average over $\l| k \ran$ states has been carried out. Physically, the basis states indices, as used in \cite{FlIz}, do not carry any significant information. However, the basis state energies $e_k(m,S)$ give the location of the corresponding strength functions and hence are meaningful \cite{Man-10}. Similarly, information entropy is computed numerically using Eq.~(\ref{eq.time19}). In order to be able to compare the fidelity and entropy results for different spins, time $t$ should have the same unit for all spins. As $\sigma_H(m,S)$ shows significant spin dependence, we express time in units of $\sigma_{avg}^{-1}$, with $\sigma^2_{avg} =[\sum_I d(m,I)]^{-1} \sum_I d(m,I) \sigma^2_H(m,I)$; $I$ is $S$ for EGOE(1+2)-$S$ and $F$ for BEGOE(1+2)-$F$.

\subsection{Fidelity}

Variation in fidelity $W_0(t)$ with time is shown in Fig.~\ref{fig.w0-f} as a function of many-particle spin $S$ for an EGOE(1+2)-$S$ ensemble. Results are shown for two choices of interaction strength: (a) $\lambda=0.3$ and (b) $\lambda=0.1$ which correspond respectively to the Gaussian and BW to Gaussian transition region for the chosen set of parameters $(\Omega,m)$. Numerical results in the figures are compared with the theoretical formulas given by Eqs. (\ref{eq.time15}) and (\ref{eq.time18}) respectively in the Gaussian and BW to Gaussian transition regimes. Agreement between numerical and theoretical EE(1+2)-$S$ results is excellent for short times. Inset to Fig.~\ref{fig.w0-f}(b) also shows the fidelity decay for longer times. Fidelity decay is slower with increasing $S$ which is in confirmation with variation of $\sigma^2_k(m,S)$ with many-particle spin for fermionic systems. The best-fit value of the parameter $\nu$ in Eq.~(\ref{eq.time18}) is found to be $\nu= 3.4$, $3$ and $2.6$ for $S=0$, $1$, and $2$ respectively and the corresponding variances $\sigma_k^2(m,S)$ are given in Table~\ref{table-sigk}.

Similarly, fidelity decay for BEGOE(1+2)-$F$ is shown in Fig. \ref{fig.w0-b}. Here, $\lambda = 0.2$ and $0.05$ correspond to Gaussian and BW to Gaussian transition regimes respectively. It is seen that agreement between numerical and analytical results is excellent. Unlike fermions, fidelity decay is faster with increasing many-particle spin $F$. This is again in confirmation with variation of $\sigma^2_k(m,F)$ with many-particle spin for bosonic systems. The best-fit value of the parameter $\nu$ in Eq.~(\ref{eq.time18}) is found to be $\nu= 5.4$, $6.6$, and $10.0$ for $F=2$, $4$, and $5$ respectively and the corresponding variances $\sigma_k^2(m,S)$ are given in Table~\ref{table-sigk}.

As the value of the interaction strength $\lambda$ increase, the eigenstates of $H$ spread over all the mean-field basis states leading to delocalization in Fock basis. Therefore, fidelity decays faster with increasing quench strength $\lambda$ both for fermionic and bosonic systems as seen from Figs.~\ref{fig.w0-f} and \ref{fig.w0-b}. Moreover,  for a given $(\Omega,m)$, bosonic systems display faster fidelity decay in comparison to fermionic systems as the dimension of the fermionic Hilbert space is smaller due to Pauli's exclusion principle. Thus, EE(1+2)-$S$ formulation captures all the essential features of time dynamics of fidelity decay as a function of many-particle spin. 

\begin{figure*}[htp] 
\begin{center} 
\includegraphics[width=0.9\linewidth]{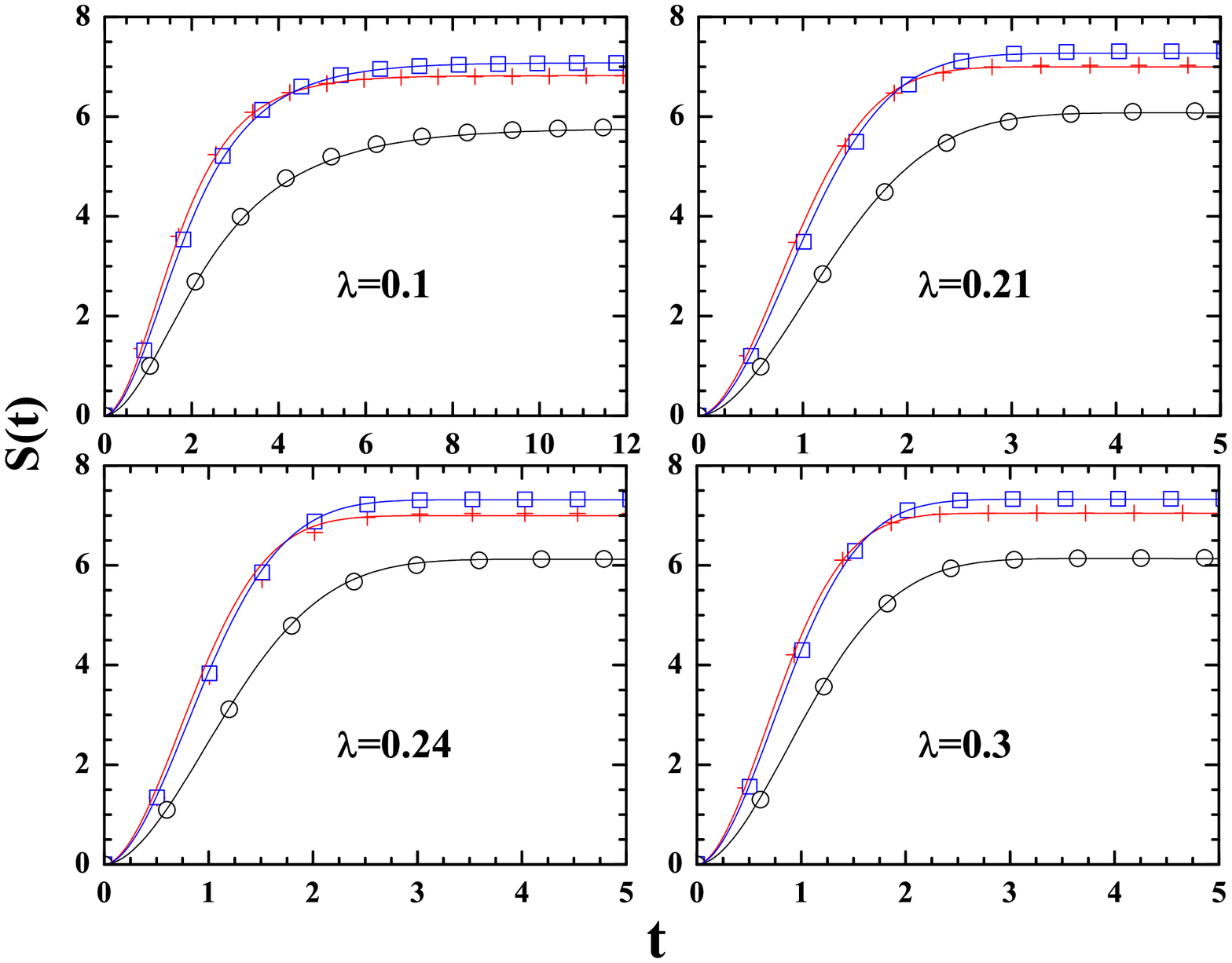}
\end{center}
\caption{Entropy $S(t)$ as function of time $t$ for different spins for
EGOE(1+2)-$S$.  The points, marked by red '+' ($S=0)$, blue 'open square'
($S=1)$ and black 'open circle' ($S=2$) correspond to numerical
results. Similarly, the continuous curves represent the corresponding 
theoretical predictions. The values of $\lambda$ are given in each
panel.}
\label{fig.ent-f}
\end{figure*}

\begin{figure*}[htp]
\begin{center}
\includegraphics[width=0.9\linewidth]{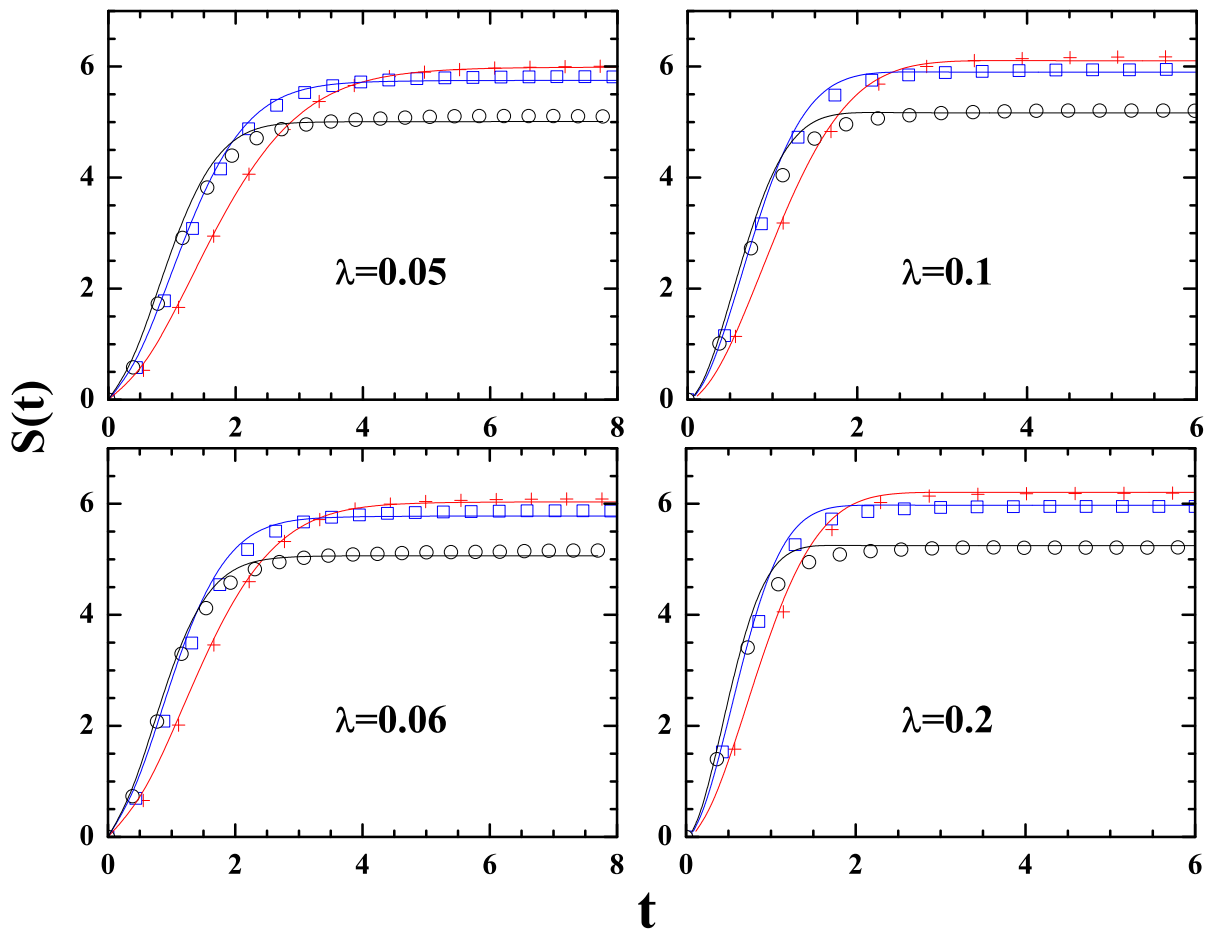} 
\end{center}
\caption{Plot of $S(t)$ as function of time $t$ for different $F$ spin values
for BEGOE(1+2)-$F$.  The points, marked by red '+' ($F=2)$, blue 'open
square' ($F=4)$ and black 'open circle' ($F=5$) correspond to numerical
results.   Similarly, the continuous curves represent the corresponding 
theoretical predictions. The values of $\lambda$ are given in each
panel.}
\label{fig.ent-b}
\end{figure*}

\begin{table}

\caption{Ensemble averaged $\sigma_k^2$ for EE(1+2)-$S$.} 
\begin{center}
\begin{tabular}{|c|c|c|c|c|c|c|c|} \hline
\multicolumn{4}{|c|}{EGOE(1+2)-$S$} & \multicolumn{4}{|c|}{BEGOE(1+2)-$F$} \\ 
\hline
 & \multicolumn{3}{|c|}{$\sigma_k^2(m,S)$} & 
& \multicolumn{3}{|c|}{$\sigma_k^2(m,F)$}\\ 
\cline{2-4}\cline{6-8}
$\lambda$ &$S=0$ & $S=1$   & $S=2 $ & $\lambda$ & $F=2$ & $F=4$   & $F=5$  \\ 
\hline
0.1    & 0.19  & 0.18  & 0.15 & 0.05 & 0.23 & 0.28 & 0.33 \\  
\hline
0.21   & 0.53   & 0.51 & 0.46 & 0.06 & 0.31 & 0.38 & 0.39 \\
\hline
0.24   & 0.59  & 0.58  & 0.53 & 0.1  & 0.56 & 0.61 & 0.61  \\
\hline
0.3    & 0.70  & 0.69  & 0.64 & 0.2  & 0.87 & 0.87 & 0.89 \\
\hline\hline
\end{tabular}
\end{center}
\label{table-sigk}
\end{table}

\subsection{Entropy}

Variation in information entropy with time as a function of many-particle spin for an EGOE(1+2)-$S$ ensemble is shown in Fig.~\ref{fig.ent-f}. We choose four values of the interaction strength: $\lambda = 0.1$ (BW to Gaussian transition region), $0.21$, $0.24$ and $0.3$ (Gaussian region). In all the cases, entropy shows a quadratic behavior for very small times,  then switches to an essentially linear growth and eventually, saturates as predicted by Eq.~(\ref{eq.theory}). Fig.~\ref{fig.ent-f} indicates a good agreement between numerical and analytical results (derived in  Secs. 3.2.1 and 3.2.3). To determine the validity of approximating $\exp[S(\infty)]$ by NPC, theoretical predictions are fitted with the numerical results for entropy. Good agreement is found between the theoretical predictions using $n$ and the numerical results with $\kappa = 2$ in the Gaussian region and $\kappa =  3$, $3.3$ and $3.6$ for $S=0-2$ respectively, in the BW to Gaussian transition region.
 
As seen from the figure, the saturation value of entropy for $S=1$ is larger compared to that for $S=2$ for a given interaction strength. This can also be understood from Eq.~(\ref{eq.theory}) as follows. Note that $n$ is a function of $\zeta^2(m,S)$ and $d(m,S)$. For a given $\lambda$, $d(m,S)$ has large variation with $S$ while $\zeta^2(m,S)$ is almost same for all spins. Therefore, the saturation value for $S=1$ is larger than for $S=2$ irrespective of $\lambda$. On the other hand, with a given spin $S$, $\zeta^2(m,S)$ increases with increasing $\lambda$ leading to enhancement in saturation value of entropy with increasing $\lambda$.

Using the calculated value of $\zeta^2(m,S)$ in Eq.~(\ref{eq.ns-th}) [Eq. (\ref{eq.npc-int})] and then applying Eq.~(\ref{eq.t-sat}) [Eq. (\ref{eq.tsat-int})], we can estimate the saturation time  $t_{sat}$ for different spins in the Gaussian [BW to Gaussian transition] region. Table~\ref{table-tsat} gives the calculated values of $t_{sat}$ for
different $S$ values. It is seen that $t_{sat}$ increases with increasing spin and this is consistent with the results in Fig.~\ref{fig.ent-f}.  

Similarly, Fig.~\ref{fig.ent-b} shows time dynamics of information entropy for a BEGOE(1+2)-$F$ ensemble as a function of many-particle spin $F$. We choose four different values of interaction strengths: $\lambda=0.1$, $0.2$ (Gaussian region) and
$\lambda=0.05$ and $0.06$ (BW to Gaussian transition region).  In all the cases, entropy shows a quadratic behavior for very small times, then switches to an essentially linear growth and eventually, saturates as predicted by Eq.~(\ref{eq.theory}). Numerical results show good agreement with analytical results derived in Secs. 3.2.1 and 3.2.3 respectively in the Gaussian and BW to Gaussian transition regimes.  Good agreement is found between the theoretical formulas and the numerical results with $\kappa = 2$ in the Gaussian region and $\kappa =  2.8$, $2.55$ and $2.3$ for $F=2$, $4$ and $5$ respectively, in the BW to Gaussian transition region. Deviations from the theoretical curves are maximum for $F=5$ for all values of $\lambda$ and this may be due to the fact that dimension of the corresponding Hilbert space is smallest compared to other spin values. Values of $t_{sat}$ are given in Table~\ref{table-tsat}. The saturation times increase with increasing many-particle spin, just as for fermionic systems.

As the parameter $\kappa$ is expected to depend on the nature of spreading of the eigenstates, it should have different values in the Gaussian and BW to Gaussian transition region. This is also in confirmation with our numerical results. In the Gaussian regime, Eqs.~(\ref{eq.theory}) and (\ref{eq.ns-th}) with $\kappa = 2$ explain the time variation in entropy with many-particle spin.  However, $\kappa$ shows dependence on many-particle spin in the BW to Gaussian transition regime. A weak dependence on spin is expected as $\zeta^2(m,S)$ varies with spin. The exact dependence of parameter $\kappa$ on $\lambda$ and many-particle spin is not explored any further in this paper. Thus, EE(1+2)-$S$ formulation captures all the essential features of time dynamics of entropy as a function of many-particle spin. 

\begin{table}
\caption{Saturation time $t_{sat}$ (in $\sigma_{avg}^{-1}$ units) for EE(1+2)-$S$. } 
\begin{center}
\begin{tabular}{|c|c|c|c|c|c|c|c|} \hline
\multicolumn{4}{|c|}{EGOE(1+2)-$S$} & \multicolumn{4}{|c|}{BEGOE(1+2)-$F$} \\ 
\hline
     & \multicolumn{3}{|c|}{$t_{sat}$} &           & \multicolumn{3}{|c|}{$t_{sat}$}\\ \cline{2-4}\cline{6-8}
$\lambda$ &$S=0$ & $S=1$ & $S=2 $  & $\lambda$ & $F=2$ & $F=4$   & $F=5$  \\ 
\hline
0.1    & 7.97  & 9.62  & 11.49 & 0.05  & 6.74  & 4.62  & 3.28 \\  
\hline
0.21   & 3.32   & 3.73 & 4.21 & 0.06 & 5.57 & 3.87 & 3.15 \\
\hline
0.24   & 3.13   & 3.5  & 3.97 & 0.1  & 3.66 & 2.68 & 2.2 \\
\hline
0.3    & 2.7    & 3.2  & 3.65 & 0.2  & 3.07 & 2.1  & 1.7 \\
\hline\hline
\end{tabular}
\end{center}
\label{table-tsat}
\end{table}

\section{Conclusions}

We have analyzed the unitary time evolution of fidelity and information entropy in an isolated interacting many-particle quantum (fermionic and bosonic) system followed by a random interaction quench as a function of many-particle spin. We obtain analytical results for time dynamics of fidelity and entropy in the framework of EE(1+2)-$S$ which are in good agreement with the numerical examples considered in the paper. We showed that the behavior of the spectral variances with spin explains the spin dependence of fidelity and entropy. There is also a dependence on the quench strength $\lambda$ which is explored in all details in the paper with respect to the three chaos markers $(\lambda_c, \lambda_\ell,\lambda_d)$ generated by EE(1+2)-$S$. It is important to note that BW to Gaussian transition in emission intensities (fidelity decay) has been studied experimentally in \cite{Ro-06} and the results are explained by the widths of the energy distributions. We have also derived approximate formulas for the saturation value $S(\infty)$ and saturation time $t_{sat}$ for the information entropy. In summary, an isolated interacting many-body fermionic (bosonic) quantum system exhibits delay in thermalization with increasing (decreasing) many-particle spin $S$($F$). Our results may contribute to formulation of a general statistical description of non-equilibrium quantum systems and in designing quantum computers.

Several approximations are used to derive the formulas for fidelity and entropy as exact formulation is non-trivial. In future, one should attempt to solve EE(1+2)-$S$ [also EE(1+2)] from the first principles. It will also be interesting to compare the formulas for $t_{sat}$ and other quantities given in this paper with the results of realistic many-body calculations such as those presented recently by Lode {\it et al.}~\cite{Axel}. Another important aspect in relaxation of quasi-integral systems is prethermalization. Recently prethermalization has been experimentally confirmed for a one dimensional degenerate Bose gas~\cite{Gring, Langen}. Therefore, it will be interesting to explore prethermalization using EE(1+2)-$S$ formalism.

The generic features explored in this paper constitute an important step towards a 
complete description of unitary time evolution of quantum systems of interacting
particles quenched far from equilibrium. This formulation gives an overall picture of relaxation of complex quantum systems in the absence of complete knowledge about system dynamics [which is generally the case for complex physical (such as nuclear, mesoscopic and atomic) systems]. Moreover, results of the present analysis may be useful in the study of the stability of a  quantum computer against quantum chaos~\cite{Shepelyansky} and to address Loschmidt echoes in many-particle quantum systems \cite{Sielgman}.

EE(1+2)-$S$ analysis of time evolution of few-body observables is in principle possible and it will be considered in a future publication. We also plan to utilize BEGOE(1+2) ensembles with spin one degrees of freedom to study non-equilibrium dynamics of isolated finite quantum systems. A method for constructing this ensemble is given in \cite{Deota-S1} although this ensemble is more challenging, computationally as well as analytically. Attempts will also be made to obtain an analytical understanding of the significance and magnitude of the parameter $\kappa$ introduced in Section 3.2.

After completing the present work, we became aware of a recent paper \cite{Mag-16}. 
Here, EGOE(1) generated by a random one-body hamiltonian (with both diagonal and
off-diagonal matrix elements) is used to show analytically that the corresponding 
fermionic systems exhibit ETH.

\begin{acknowledgments}

Thanks are due to Barnali Chakrabarti and Luis Benet for many useful discussions. NDC acknowledges financial support from UGC, India under a research project [Grant No. F.40-425/2011 (SR)]. MV acknowledges supercomputing facility SC15-1-IR-61-MZ0 (DGTIC-UNAM) and financial support from UNAM/DGAPA/PAPIIT research grant IG100616 and CONACyT research grant 219993. Most of the numerical calculations were done using Physical Research Laboratory's VIKRAM-100 HPC cluster.

\end{acknowledgments}

\begin{widetext}
\section*{Appendix}

We give the formulas for the variance propagators $P(\Omega,m,S)$ and $Q(\Omega,m,S)$ for EGOE(1+2)-$S$ and BEGOE(1+2)-$F$ derived in \cite{Ma-PRE} and \cite{Ma-F} respectively using trace propagation method:
\be
\barr{l}
P(\Omega,m,S) = 
\dis\frac{1}{\Omega(\Omega+1)/2}
\l[\dis\frac{\Omega+2}{\Omega+1} \cq^1(\{2\}:m,S) +
\dis\frac{\Omega^2+3\Omega+2}{\Omega^2+3\Omega}\,\cq^2(\{2\}:m,S)\r] \\
 + \dis\frac{1}{\Omega(\Omega-1)/2}
\l[\dis\frac{\Omega+2}{\Omega+1} \cq^1(\{1^2\}:m,S) +
\dis\frac{\Omega^2+\Omega+2}{\Omega^2+\Omega}\,\cq^2(\{1^2\}:m,S)\r]\;;\\
\cq^1(\{2\}:m,S) = \l[(\Omega+1) \capp^0(m,S)/16\r]\,\l[m^x(m+2)/2 + 
\cas^2 \r]\,,\\
\cq^2(\{2\}:m,S) = \l[\Omega (\Omega+3) \capp^0(m,S)/32\r]\,\l[m^x(m^x+1) -
\cas^2 \r]\,,\\
\cq^1(\{1^2\}:m,S) = \dis\frac{(\Omega-1)}{16(\Omega-2)}
\l[(\Omega+2)\,\capp^1(m,S)
\,\capp^2(m,S) \r. \\
+ \l. 8\Omega (m-1)(\Omega-2m+4) \cas^2 \r]\;, \\
\cq^2(\{1^2\}:m,S)  = \dis\frac{\Omega}{8(\Omega-2)} 
\l.[(3\Omega^2 -7\Omega +6)
(\cas^2)^2 \r. \\
 + 3m(m-2)m^x(m^x-1)(\Omega+1)(\Omega+2)/4 \\
 + \l. \cas^2 \l\{-m m^x (5\Omega-3)(\Omega+2)+
\Omega(\Omega-1)(\Omega+1)(\Omega+6)\r\}\r]\;,\\
\capp^0(m,S) = \l[ m(m+2) - 4\cas^2\r]\;,\;\;\;
\capp^1(m,S) = \l[3m(m-2) + 4\cas^2\r]\,, \\ 
\capp^2(m,S) = 3 m^x(m-2)/2 - \cas^2 \;,\;\;\;
m^x=\l(\Omega-\dis\frac{m}{2}\r)\;,\;\;\;\cas^2=S(S+1)\;.
\earr
\label{appndx-1}
\ee
\be
\barr{l}
Q(\Omega,m,F) = \dis\sum_{f=0,1}
(\Omega-1)(\Omega-2(-1)^f)(\Omega+2)\;\car^{\nu=1,f}(m,F) \\
+ \dis\frac{(\Omega-3)(\Omega^2+\Omega+2)}{2(\Omega-1)}
\; \car^{\nu=2,f=0}(m,F)
+ \dis\frac{(\Omega-1)(\Omega+2)}{2}\; \car^{\nu=2,f=1}(m,F)\;; \\
\car^{\nu=1,f=0}(m,F) = \dis\frac{\l[(m+2)m^\star/2 - \lan F^2 \ran\r]
\cx^0(m,F)}
{8 (\Omega-2) (\Omega-1) \Omega (\Omega+1) } \;,\\
\car^{\nu=1,f=1}(m,F) =\dis\frac{8\Omega(m-1)(\Omega+2m-4) \lan F^2 \ran +
(\Omega-2) \cx^2(m,F) \cx^1(m,F)}
{8  (\Omega-1) \Omega (\Omega+1) (\Omega+2)^2} \;,\\
\car^{\nu=2,f=0}(m,F) = \l[m^\star(m^\star-1) - \lan F^2 \ran\r]
\cx^0(m,F)/[8 \Omega (\Omega+1)]\;,\\
\\
\car^{\nu=2,f=1}(m,F) = \l\{\l[\lan F^2 \ran\r]^2 (3\Omega^2+7\Omega+6)/2 +
3m(m-2) m^\star (m^\star+1) \times \r. \\
\l. (\Omega-1)(\Omega-2)/8 + \l[\lan F^2 \ran/2\r]\l[(5\Omega+3)
(\Omega-2)m m^\star +
\Omega(\Omega-1)(\Omega+1)(\Omega-6)\r]\r.\}/ \\
\l[(\Omega-1) \Omega (\Omega+2)(\Omega+3)\r]\;; \\
\cx^0(m,F) = \l[ m(m+2) - 4 \lan F^2 \ran \r]\;,\;\;\;
\cx^1(m,F) =  \l[3m(m-2) + 4 \lan F^2 \ran \r]\;,\\ 
\cx^2(m,F) = 3(m-2) m^\star/2 + \lan F^2 \ran\;,\;\;\;\;m^\star =
\Omega+m/2\;,\;\;\;\; \lan F^2 \ran =F(F+1)\;.
\earr
\label{appndx-2}
\ee
\end{widetext}

\end{document}